\documentclass{article}

\usepackage[preprint]{kobi}
\usepackage{graphicx}
\usepackage{caption}
\usepackage{float}
\usepackage{lmodern}
\usepackage[utf8]{inputenc}
\usepackage[T1]{fontenc}
\usepackage{xcolor}
\definecolor{darkblue}{rgb}{0.1, 0.2, 0.75}
\usepackage[hidelinks]{hyperref}
\hypersetup{
    colorlinks=true,
    linkcolor=darkblue,
    citecolor=darkblue,
    urlcolor=darkblue
}
\usepackage{url}
\usepackage{booktabs}
\usepackage{amsfonts}
\usepackage{nicefrac}
\usepackage{microtype}
\usepackage{lineno}
\usepackage{tabularx}
\usepackage{makecell}
\usepackage{threeparttable}
\usepackage{amsmath}
\usepackage{footnote}

\title{Artificial intelligence can persuade people to take political actions}

\author{
  Kobi Hackenburg$^{1,2}$\thanks{K.H. and L.H. contributed equally to this work. To whom correspondence should be addressed. E-mail: \href{mailto:kobi.hackenburg@oii.ox.ac.uk}{\texttt{kobi.hackenburg@oii.ox.ac.uk}}, \href{mailto:lbh@stanford.edu}{\texttt{lbh@stanford.edu}}} ,
  Luke Hewitt$^{3,2}$\footnotemark[1] ,
  Caroline Wagner$^{2}$, \\
  \textbf{Christopher Summerfield}$^{1,2}$,
  \textbf{Ben M. Tappin}$^{4}$ \and
  \\
{$^1$University of Oxford, Oxford, OX2 6GG, UK}\\
{$^2$UK AI Security Institute, London, UK}\\
{$^3$Stanford University, Stanford, CA 94305, USA}\\
{$^4$London School of Economics and Political Science, London, WC2A 2AE, UK}
}

\begin{document}

\maketitle
\vskip 0.2in
\begin{abstract}
\leftskip=.25in
\rightskip=.25in
There is substantial concern about the ability of advanced artificial intelligence to influence people's behaviour. A rapidly growing body of research has found that AI can produce large persuasive effects on people's attitudes, but whether AI can persuade people to take consequential real-world actions has remained unclear. In two large preregistered experiments (N=17{,}950 responses from 14{,}779 people), we used conversational AI models to persuade participants on a range of attitudinal and behavioural outcomes, including signing real petitions and donating money to charity. We found sizable AI persuasion effects on these behavioural outcomes (e.g. +19.7 percentage points on petition signing). However, we observed no evidence of a correlation between AI persuasion effects on attitudes and behaviour. Moreover, we replicated prior findings that information provision drove effects on attitudes, but found no such evidence for our behavioural outcomes. In a test of eight behavioural persuasion strategies, all outperformed the most effective attitudinal persuasion strategy, but differences among the eight were small. Taken together, these results suggest that previous findings relying on attitudinal outcomes may generalize poorly to behaviour, and therefore risk substantially mischaracterizing the real-world behavioural impact of AI persuasion.

\vskip 0.4in

\end{abstract}


\newpage
\section*{Introduction}
\label{sec:intro}

There is substantial concern about the ability of artificial intelligence to influence people's behaviour \cite{luciano2024hypersuasion,burtell2023artificial,chen2025framework,elsayed2024mechanism,summerfield2024democracy}, and a rapidly growing body of research has studied the persuasive influence of AI in the political domain \cite{chen2025framework,hackenburg2025levers,goldstein2024persuasive,bai2023artificial,costello2024durably,hackenburg2025scaling,schoenegger2025llms,simchon2024persuasive,argyle2025testing}. However, existing evidence has been almost entirely limited to measuring the effects of AI on people's attitudes rather than their actual behaviour. Given research in other domains showing that treatment effects on people's attitudes and on their behaviour do not always correspond \cite{hainmueller2015validating,okeefe2021persuasive,coppock2015assessing,saccardo2024field}, considerable uncertainty remains about three key questions regarding persuasion by advanced AI.

First, \textit{how large are AI persuasion effects on behavioural outcomes?} This question is important for understanding the real-world scope of AI influence: prior work has reported AI persuasion effects of 10 percentage points (pp) or more on attitude outcomes \cite{hackenburg2025levers,goldstein2024persuasive,costello2024durably}; finding comparable effects on behavioural outcomes would bolster the claim that AI persuasion could have substantial real-world impact. Second, \textit{are AI persuasion effects on attitude outcomes correlated with those on behavioural outcomes?} This question is important for understanding what we can learn from research that uses only attitude outcomes: if there is little correlation, then findings from prior work comparing different treatment conditions---such as conversation vs.\ static messaging, or laypeople vs.\ AI---may generalize poorly to behaviour, which is often the ultimate outcome of interest. Third, \textit{is the mechanism for AI persuasion on behavioural outcomes similar to that for attitude outcomes?} Prior work suggests that a key mechanism of AI persuasion on attitude outcomes is packing conversations with factual claims \cite{hackenburg2025levers,costello2024durably}; is this also the case for behavioural outcomes?

\begin{figure}[H]
\centering
\includegraphics[width=\linewidth]{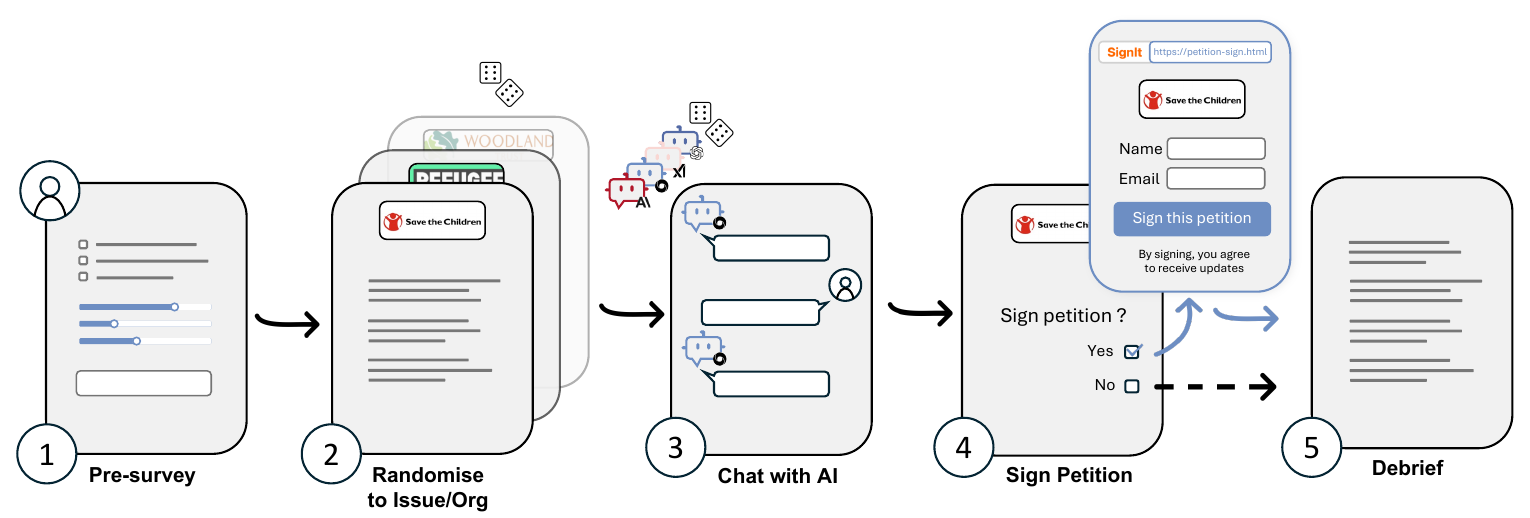}
\caption{\textbf{Experimental procedure for measuring real-world petition-signing.} Participants provided pretreatment measures (Step~1), were randomly assigned to one of eight real UK petitions and sponsoring organisations, (Step~2; see \textit{Methods} Table \ref{tab:petitions}), and engaged in a multi-turn conversation with a frontier AI model (Step~3; see \textit{Methods} for list of models) instructed to persuade them to sign the petition. After the conversation, participants were asked whether they wished to sign (Step~4). Those who indicated willingness could optionally navigate to a separate website, where they could sign the petition by entering their first name, last name, and email address, agreeing to receive email updates from the organization, and clicking ``Sign this petition''. Participants could skip or abandon this process at any point, and only those who completed all steps were counted as having signed. All participants were subsequently debriefed (Step~5). For screenshots of the petition-signing website, see SI Appendix Figures S10-11. Additional outcome measures varied by study; see \textit{Methods}.}
\label{fig:methods}
\end{figure}

Here, we answer these questions through large AI-human persuasion experiments in which we measure both attitude outcomes and behavioural outcomes on a range of political topics. Across two studies, we recruited 14{,}779 UK adults (17{,}950 total study responses) and deployed frontier AI models (\textit{e.g.}, GPT-4.1, Claude Opus 4.6, Grok 4, Gemini 3.1 Pro) to persuade them via interactive conversation to take political action on one of eight different causes---including signing a real petition using their name and email address (see Figure \ref{fig:methods}), and donating money to the sponsoring organization. The petitions were diverse, spanning topics such as nuclear disarmament, democratic reform, and animal welfare. We also measured people's attitudes toward the petition and the sponsoring organization on 7-point Likert scales. We further tested eight theory-driven persuasion strategies to investigate what mechanisms drive AI persuasion on behavioural outcomes. Conversations lasted an average of 4.9 turns (7 minutes).

\section*{Results}

\subsection*{AI conversations produce sizable persuasion effects on behaviour}

\begin{figure}[tbp]
\centering
\includegraphics[width=0.6\linewidth]{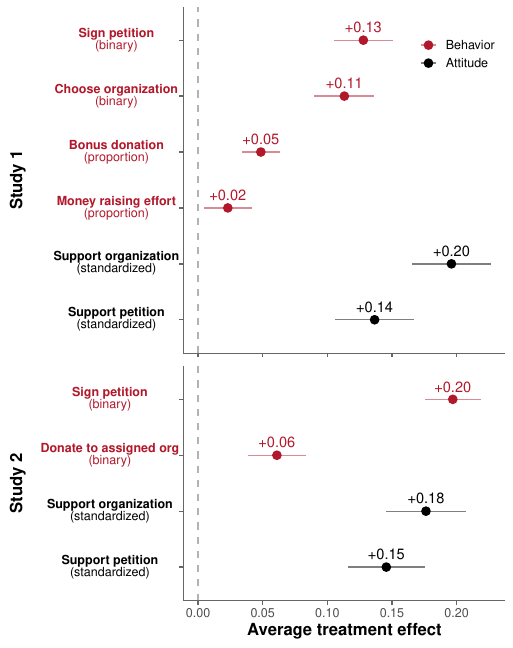}
\caption{\textbf{AI conversations produce sizable persuasion effects on political behaviours.} Average treatment effects (ATEs) of AI conversation vs.\ control on behavioural outcomes (red) and attitude outcomes (black), with 95\% confidence intervals. \textbf{Study~1} ($N=8{,}000$): four behavioural outcomes (petition signing, choosing to keep the donation with the petition-sponsoring organization vs.\ switching to an alternative charity, bonus donation, and money-raising effort via a clicking task \cite{dellavigna2018motivates}) and two attitude outcomes (petition support and organization support, measured as standardized pre--post change). \textbf{Study~2} ($N=9{,}950$):  two behavioural outcomes (petition signing and choosing to donate to the assigned organization vs.\ an alternative) and two attitude outcomes (petition support and organization support) .}
\label{fig:main-effects}
\end{figure}

\begin{figure}[tbp]
\centering
\includegraphics[width=\linewidth]{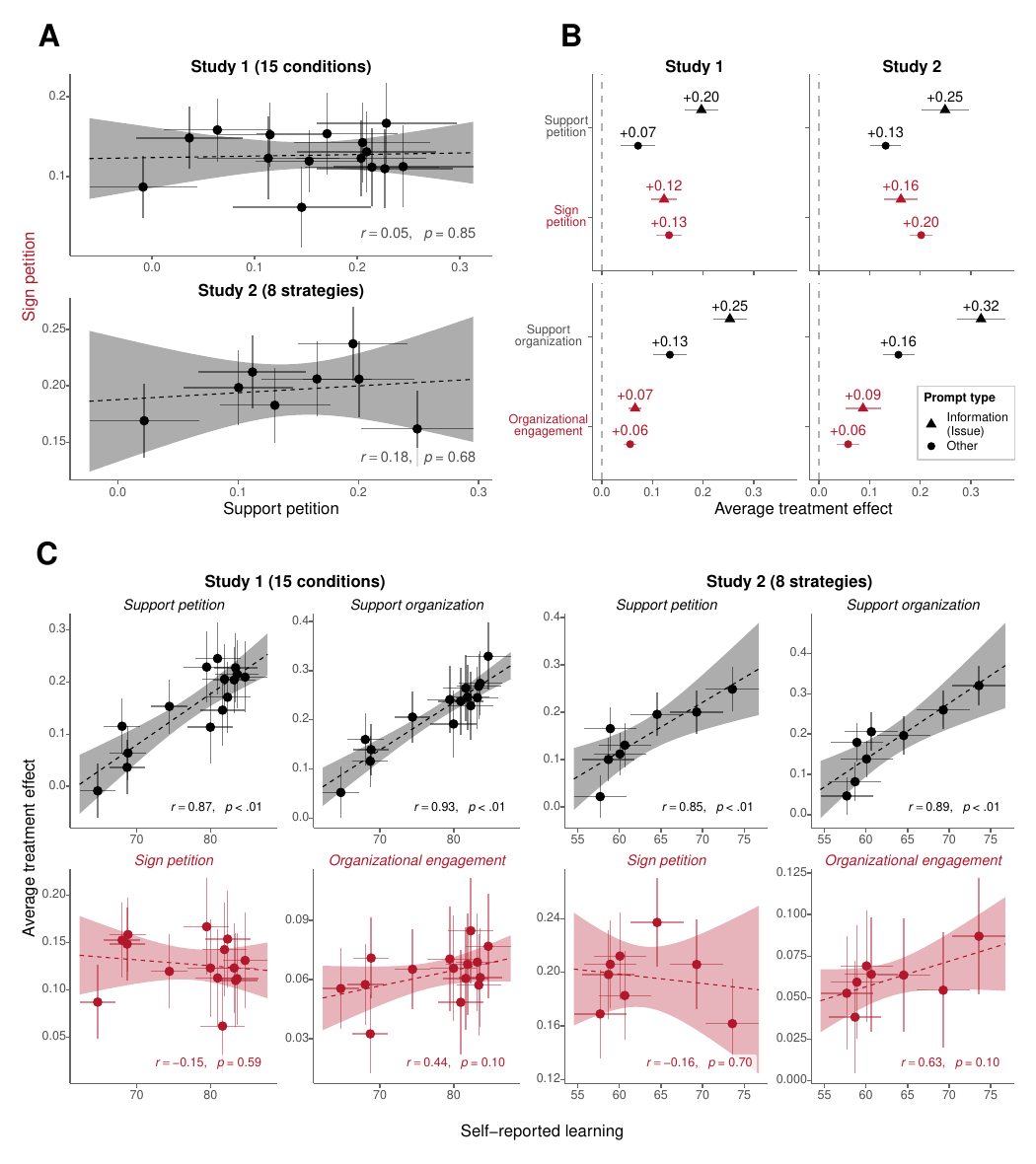}
\caption{\textbf{AI persuasion effects on attitudes and behaviour are uncorrelated and driven by different mechanisms.} \textbf{(A)}~Treatment effect on petition support (x-axis) vs.\ treatment effect on petition signing (y-axis). Each point represents one experimental condition; dashed line shows the linear fit with 95\% confidence band. Study~1: 15 conditions (5 AI models $\times$ 3 conversation types). Study~2: 8 persuasion strategies. No significant correlation is observed in either study. \textbf{(B)}~Average treatment effects on attitude and behavioural outcomes, split by prompt type: ``Information (Issue)'' vs.\ ``Other.'' In Study~1, ``Other'' refers to the Random prompt condition, in which models received one of 200 randomly generated persuasion prompts. In Study~2, ``Other'' refers to the pooled average of the seven non-Info-Issue strategies (Emotional Activation, Implementation Intentions, Identity Labeling, Commitment Escalation, Anticipated Regret, Info: Impact Efficacy, and Mega). Information prompts produce relatively larger effects on attitude outcomes (top rows) but not on behavioural outcomes (bottom rows). \textbf{(C)}~Self-reported learning (x-axis; participants' agreement that they learned new information from the conversation, measured on a 0–100 scale) vs.\ treatment effect on each outcome (y-axis). Each point represents one condition/strategy. Learning is strongly associated with attitude persuasion (top row) but not behavioural persuasion (bottom row), in both studies.}
\label{fig:mechanisms}
\end{figure}

Fig.~\ref{fig:main-effects} shows the estimated persuasive effect of a persuasive AI conversation about the assigned petition (vs.\ a control group that engaged in an AI conversation on a neutral, non-political topic) on all outcomes (SI Appendix, Tables S8-S9). In Study~1, we find large effects on petition signing: participants in the treatment group were $12.8$pp more likely to sign the petition ($p<.001$). Participants also donated \pounds0.08 more to the sponsoring organization (+ 4.9pp of bonus donated, $p < .001$, or \pounds0.12 in the larger \pounds2--4 bonus conditions, $p < 0.001$). We included two further behavioural outcomes related to the organization: engagement in an effortful task (clicking the screen up to 100 times) to generate more donations \cite{dellavigna2018motivates}, and deciding whether or not to switch the donation to an alternative charity at the end of the survey. Participants in the treatment group made more clicks in the effortful task ($+2.3$ clicks, $p = .015$) and were far more likely to stick with donating to the petition-sponsoring organization rather than switch ($+11.3$pp, $p < .001$). These results are fairly consistent across the eight distinct petitions and corresponding organizations in our sample (see Methods, Table~\ref{tab:petitions}; SI Appendix, Table S8, Figure S2). However, we note that the effect on the effortful clicking task is partly driven by the ``Stop Rising Child Poverty'' petition, and is no longer statistically significant when that issue is excluded (SI Appendix, Section 2.8.2).

\begin{figure}[tbp]
\centering
\includegraphics[width=\linewidth]{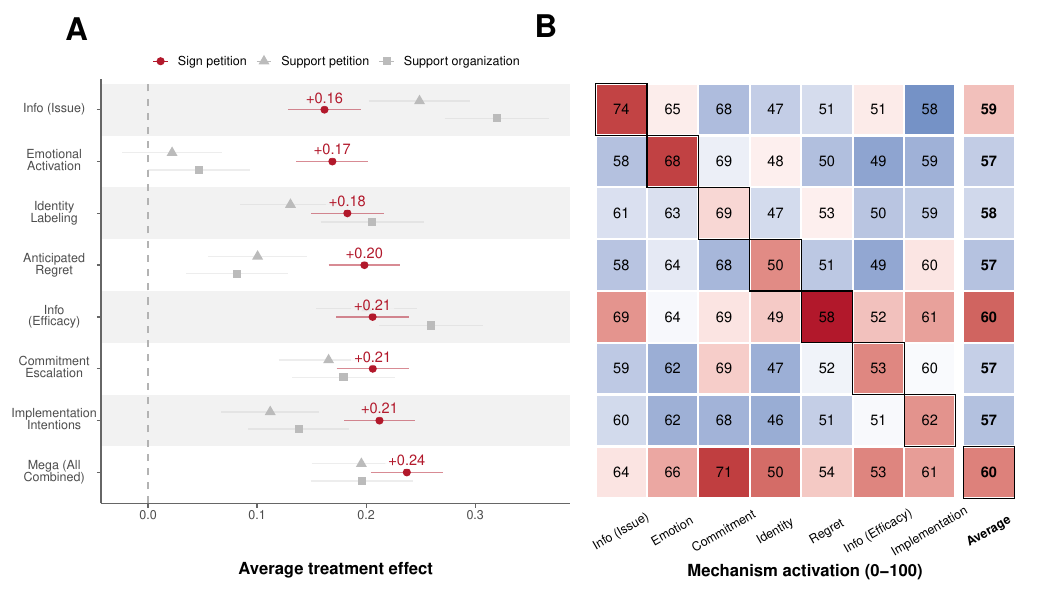}
\caption{\textbf{A combined persuasion strategy is most effective at driving behavioural change.} \textbf{(A)}~Average treatment effects (vs.\ control) of each Study~2 persuasion strategy on petition signing (red circles), petition support (grey triangles), and organization support (grey squares). Strategies are ordered by their effect on petition signing. Error bars represent 95\% confidence intervals. ``Mega (All Combined)'' is a combined strategy that integrates all other strategies into a single prompt; the LLM adaptively selects which persuasion techniques to deploy during the conversation. \textbf{(B)}~Heatmap of mean self-reported mechanism activation (0--100 scale) for each strategy (rows) across seven mechanism batteries (columns), each corresponding to the theoretical basis of one strategy. Cell values show mean scores; colour indicates relative activation within each mechanism (z-scored by column; red = above average, blue = below average). Black-outlined cells mark the diagonal---where a strategy is paired with its own targeted mechanism. The rightmost column (bold) shows each strategy's average activation across all mechanisms. The Mega strategy shows elevated activation across most mechanisms, consistent with its superior persuasiveness on behavioural outcomes.}
\label{fig:strategy-effects}
\end{figure}

In Study~2, we replicate these findings with a new sample and new set of persuasion strategies (SI Appendix, Table S9, Figure S3). Treatment participants were $19.7$pp more likely to sign the petition and $6.1$pp more likely to allocate their charity earnings to the petition-sponsoring organization rather than an alternative (both $p < .001$). Effects on attitude outcomes were also significant ($+0.15$ standardized units for petition support; $+0.18$ for organization support). These results are robust to the inclusion of demographic covariates (age, gender, education, and income), a leave-one-issue-out procedure, alternative model specifications, and were broadly consistent across AI models (SI Appendix, Section 2.8 and Figure S4).

To contextualise these effect sizes within the broader literature, we conducted a review of randomised experiments measuring petition-signing outcomes (SI Appendix, Figure S9). Our effects are comparable to, and in many cases larger than, those observed across survey experiments and real-world field studies using other treatment modalities, including face-to-face canvassing and direct messaging campaigns (see Discussion).

In summary, across two studies we find sizable AI persuasion effects on behaviours, including over $10$ percentage points on signing a real-world petition about topics such as democratic reform and nuclear disarmament. This suggests that even brief conversations with current LLMs could meaningfully influence politically relevant behavior. Though we note that such engagement may be difficult to scale outside of a paid survey \cite{chen2025framework,hackenburg2025levers}.

\subsection*{Attitude change and behaviour change are uncorrelated and driven by different mechanisms}

Having established that AI conversations can shift behaviour, an important question is whether persuasion effects on attitudes and behaviour are related. If not, then findings from the large body of prior work comparing AI persuasion approaches on attitude outcomes may generalize poorly to behaviour.

We examine this question by comparing the estimated persuasive effects of AI conversation on attitude outcomes vs.\ behavioural outcomes across each of the 15 randomized conditions in our Study~1 design (5 AI models crossed with 3 conversation conditions; see \textit{Methods}). We do not find evidence of an association: there was no significant correlation between effects on petition support and petition signing ($r=0.05, p=0.85$; Fig.~\ref{fig:mechanisms}A, top). In Study~2, we replicate this null result across the eight persuasion strategies ($r=0.18, p=0.68$; Fig.~\ref{fig:mechanisms}A, bottom). The conditions that were most persuasive on attitudes were not the most persuasive on behaviour.

This dissociation suggests that AI persuasion on attitudes and behaviour may operate via different mechanisms. Prior work provides converging evidence that a key mechanism of AI persuasion on attitude outcomes is packing conversations with factual claims \cite{hackenburg2025levers,costello2024durably}.  We directly test this mechanism here in two ways.

First, across the 15 randomized conditions in Study~1, participants' post-treatment self-reports of how much they learned from the conversation were strongly correlated with treatment effects on attitudes ($r=0.87, p<.01$ for petition support; Fig.~\ref{fig:mechanisms}C, top row) but not with effects on behaviour ($r=-0.15, p=0.59$ for petition signing; Fig.~\ref{fig:mechanisms}C, bottom row). In Study~2, this pattern replicates: self-reported learning predicts attitude change ($r=0.85, p<.01$) but not behavioural change ($r=-0.16, p=0.70$).

Second, prompting models to use information resulted in greater persuasion on attitude outcomes than other prompts, but not on behavioural outcomes (Fig.~\ref{fig:mechanisms}B). In Study~1, the ``Information (Issue)" prompt produced an effect of $+0.20$ on petition support compared to $+0.07$ for other prompts, but effects on petition signing were comparable ($+0.12$ vs.\ $+0.13$). In Study~2, the ``Information (Issue)" strategy similarly ranked among the most effective for attitude change but was the \textit{least} effective for behavioural change ($+0.16$ on petition signing vs.\ $+0.20$ for other strategies; Fig.~\ref{fig:mechanisms}B). (We note that the Information (Issue) prompting strategy we use here has been shown to massively increase the density of verifiable factual claims in AI-generated conversations \cite{hackenburg2025levers}.)

In summary, we find converging evidence that information provision drives AI persuasion on attitudes but not on behaviour. This raises a critical question: if not information, what \textit{does} drive behavioural persuasion?

\subsection*{What drives behavioural persuasion?}

To investigate this question, Study~2 tested eight theory-driven persuasion-to-action strategies head-to-head: Emotional Activation, Implementation Intentions, Identity Labeling, Commitment Escalation, Anticipated Regret, Information (Issue) (the information prompt from Study~1, also found to be most effective in prior work \cite{hackenburg2025levers}), Information---Impact Efficacy, and a ``Mega'' strategy that integrates all of the above and allows the model to adaptively deploy them during the conversation (see \textit{Methods}). These strategies were sourced from existing literature to capture the most prominent mechanisms of persuasion.

All eight strategies produced significant positive effects on petition signing relative to the control condition (Fig.~\ref{fig:strategy-effects}A; all $p < .001$ after Benjamini--Hochberg correction). However, the strategies differed in their behavioural effectiveness (Wald $\chi^2(7) = 22.42$, $p = .002$). The Mega strategy was the most effective, increasing petition signing by $23.7$pp, followed by Implementation Intentions ($+21.2$pp) and Commitment Escalation ($+20.6$pp). The Information---Issue strategy, which was the most effective on attitude outcomes, was the \textit{least} effective on behaviour ($+16.2$pp)---further confirming that information provision drives attitude change but not behavioural change.

That the Mega strategy performed best suggests  a combined strategy outperforms any single rhetorical approach. This idea is supported by the mechanism heatmap analysis in Fig.~\ref{fig:strategy-effects}B. After their conversation, we asked participants on a 0-100 scale how strongly they experienced each of the theory-driven psychological mechanisms that the strategy prompt was designed to activate (e.g., the Emotion battery for Emotional Activation, the Implementation battery for Implementation Intentions). The heatmap presents these ratings and reveals two key patterns. First, individual strategies were broadly successful in activating their targeted mechanism: diagonal cells (where a strategy is paired with its own mechanism) are systematically the highest-scoring cells in their row. Second, the Mega strategy shows elevated activation across nearly all mechanisms---its average activation score (bold column) is among the highest of any strategy---consistent with the idea that it adaptively deploys multiple persuasion techniques within a single conversation. Together, these results suggest that behavioural persuasion, unlike attitudinal persuasion, may not depend centrally on any single persuasion strategy: all eight strategies produced sizable behavioural effects within a relatively narrow range. The Mega strategy---which could draw on all available pathways---was the most effective, but its advantage over the best single-strategy conditions was modest.

\section*{Discussion}

A rapidly growing body of research has studied the persuasive effects of  AI systems \cite{chen2025framework,hackenburg2025levers,goldstein2024persuasive,bai2023artificial,costello2024durably,hackenburg2025scaling,schoenegger2025llms,simchon2024persuasive,argyle2025testing,akbulut2026gdm}, motivated by concerns that advanced AI will be used to influence people's behaviour \cite{luciano2024hypersuasion,burtell2023artificial,chen2025framework,elsayed2024mechanism,summerfield2024democracy}. However, existing work overwhelmingly measures people's attitudes, not their behaviour, and treatment effects on these outcomes do not always correspond \cite{hainmueller2015validating,okeefe2021persuasive,coppock2015assessing,saccardo2024field}. This leaves much uncertainty over key questions related to AI's persuasive influence.

Here we reported three key results. First, we find clear and sizable AI persuasion effects on behavioural outcomes, including $>$10pp on signing a real-world petition about topics such as democratic reform and nuclear disarmament. A review of randomised experiments measuring petition-signing outcomes (SI Appendix, Figure S9) indicates that these effect sizes are comparable to, and in many cases larger than, those reported in survey experiments and real-world field studies employing other treatment modalities---including face-to-face canvassing \cite{Paler2018socialcosts} and direct messaging campaigns \cite{Coppock2016field}---that social scientists already treat as practically meaningful. (We note, however, that heterogeneity across the included studies is considerable and that conversational AI represents a distinct treatment modality from those tested in prior work.) These results suggest that, if actors can get people to engage with current frontier AI systems for a brief conversation ($<$10 min), it could meaningfully influence their behaviour in a politically relevant sense.

Second, we observe no evidence of a correlation between AI persuasion effects on attitudes and effects on behaviour. Thus, insofar as studies of AI persuasion aim to learn about the effects of different factors (e.g., LLM vs. layperson, different prompts, personalized vs. generic messaging, etc.) on people's behaviour, our results suggest they urgently need to measure behavioural outcomes—otherwise their conclusions may be misleading.

Third, and perhaps explaining our second result, the mechanism underlying AI persuasion effects on behaviour does not appear to be the same mechanism as for effects on attitudes. We clearly replicated the evidence for the mechanism on attitudes reported in previous work---information \cite{hackenburg2025levers, costello2024durably}---but did not see any evidence of this mechanism for behaviour. In Study~2, we tested eight theory-driven behavioural persuasion strategies head-to-head to investigate what does drive behavioural change. A combined ``Mega'' strategy---which could flexibly select among and combine the individual strategies---was most effective, but it is notable how little the strategies varied on behaviour compared to attitudes: effects on petition signing spanned a $7.5$pp range ($+16.2$ to $+23.7$pp), whereas effects on petition support spanned an order of magnitude ($+0.02$ to $+0.25$ standardized units), with some strategies failing to produce significant attitude change at all. This implies that the choice of persuasion strategy could matter considerably more for attitudinal outcomes than for the behavioural outcomes we test here. It also suggests that, rather than selecting the right rhetorical technique, the primary bottleneck for AI-powered behavioural persuasion may be obtaining exposure or engagement \cite{tappin2025exposure}.

The attitude-behaviour mechanism dissociation we observe could be understood through the well-documented distinction between persuasion and mobilization in politics \cite{gerber2008social}: information can change people's minds, but effectively motivating costly behavioural action may require different techniques. Consistent with this interpretation, in additional exploratory analyses, we find that treatment effects on attitudes were concentrated among non-supporters (persuasion of the unconvinced), whereas treatment effects on behaviour were concentrated among prior supporters (mobilization of the already sympathetic; SI Appendix, Figures S5-S6). A robustness check indicates that this dissociation is not attributable to measurement ceiling effects (SI Appendix, Sections 2.6 and 2.8.1). This result carries practical implications: actors seeking to use AI to change people's minds face a different challenge than those seeking to mobilize action among the already sympathetic, and the effectiveness of a given AI persuasion approach may depend critically on which goal is being pursued. More broadly, it underscores that studies measuring only attitudinal outcomes may systematically overlook AI's capacity to mobilize behaviour among those who already agree but have not yet acted---a population that is arguably of greatest concern for real-world political influence.

We note several limitations. First, our studies were conducted in a paid survey context in which participants were paid to engage with an AI system, providing a degree of captive attention that may not characterise naturalistic encounters with AI-generated persuasion. We therefore note it may be difficult to scale such engagement outside of a paid survey \cite{chen2025framework, hackenburg2025levers, tappin2025exposure}. Second, the behaviours we measure, while consequential, are relatively low-cost, and future research should investigate whether these effects would generalize to higher-stakes behaviours such as vaccination, news consumption or dieting.

Taken together, our findings indicate that the rapidly growing body of research measuring AI persuasion on attitudes may be building an incomplete---and potentially misleading---picture of AI's real-world influence. Changing someone's mind and getting them to act appear to be distinct processes, driven by different mechanisms and affecting different populations. Understanding and governing the political impact of advanced AI will require research that takes behaviour, not just attitudes, as the outcome of interest.

\section{Methods}
\label{sec:methods}

The present studies were approved by the Research Assurance Board at the UK AI Security Institute and carried out in accordance with the ethical principles of the Declaration of Helsinki. Informed consent was obtained from all participants. Study 1 was preregistered on \href{https://aspredicted.org/h4yg27.pdf}{AsPredicted}; Study 2 on the \href{https://osf.io/bf7q4/overview}{Open Science Framework}. Study 1's preregistered analyses comprised the primary average treatment effects on all six outcomes. Analyses of the attitude–behaviour correlation, the role of information provision, and the association between self-reported learning and treatment effects were exploratory in Study 1 but preregistered for Study 2.

\subsection{Sample}
Across both studies, participants were recruited via Prolific, an online recruitment platform shown to yield high-quality data \cite{peereyal2022prolific, stagnaro2025prolific}. Participants were screened to include UK-residents aged 18 or older who were fluent in English. The compensation rate was approximately £12 per hour and participants were debriefed upon completion (SI Appendix, Section 3.7). Participants who failed a Cloudflare Turnstile bot verification or an attention check were excluded, with list-wise deletion for remaining missing data. Additionally, some participants dropped out after treatment assignment but before providing their outcome variable, resulting in overall post-treatment attrition rates of 4.0\% (Study 1) and 2.9\% (Study 2). There was no evidence of differential attrition across either model or persuasion strategy conditions. For sample composition, attrition rates, and robustness checks, see SI Appendix, Sections 2.1, 2.3, 2.7 and 2.8.

\textit{Study 1.} Data were collected between September 18–23, 2025. In addition to their base compensation, participants received a study bonus (randomized to either £0.50, £1, £2, or £4 for generalization). The preregistered target sample was 8{,}000 participants. A total of 8{,}412 participants were randomized. Of these, 339 (4.0\%) did not complete the study after treatment assignment and 73 failed the attention check, yielding a final analytical sample of 8{,}000 participants.

\textit{Study 2.} Data were collected between March 5--17, 2026. The preregistered target sample was approximately 10,000 participants (with 1,000 for each of the eight treatment conditions and 2,000 in the control condition). A total of 10{,}317 participants were randomized. Of these, 301 (2.9\%) did not complete the study after treatment assignment and 66 failed the attention check, yielding a final analytical sample of 9{,}950 participants.

\subsection{Experimental design}
Both studies employed between-subjects designs in which participants were randomly assigned to a conversation condition (\~20\% control; remainder equal across treatments) and to one of eight political issues, each associated with a real UK petition and sponsoring organization (Table~\ref{tab:petitions}; SI Appendix, Tables S2-S5 for participant counts per condition; Tables S6-S7 for pre-treatment attitudes and control-group outcome levels by petition topic).

Participants completed demographic questions and pretreatment measures of petition and organization support (7-point Likert scales), engaged in a multi-turn text conversation with a frontier LLM (2–10 turns), and then again reported their petition and organization support. Finally, respondents signed the real petition through a pop-up interface embedded in the study, entering their name and email address, which were submitted to the petition website without being stored by the researchers. They were also given the opportunity to direct funds to the sponsoring organization or to GiveWell, an independent charity evaluator (for procedural details and scale items, see SI Appendix, Section 3).

\textit{Study 1.} Participants were randomised across issue (8 levels), AI model (GPT-4o, GPT-4.1, Claude Sonnet 4, Grok 4, Gemini 2.5 Pro), study bonus (£0.50, £1, £2, or £4), and conversation type. Conversation type varied the system prompt delivered to the LLM: a neutral control conversation about an unrelated topic, two fixed information-based persuasion prompts targeting the petition or the sponsoring organization, or a prompt randomly sampled from a pool of 200 (SI Appendix, Section 3.8.3). Mean conversation length was 4.9 turns. Beyond petition signing, three additional behavioural outcomes were collected: donation of any portion of the study bonus, a repetitive clicking task generating £0.03 per click (up to £3) for the organization, and whether participants reallocated raised funds to GiveWell.

\textit{Study 2.} To investigate what drives behavioural persuasion, Study~2 tested eight theory-driven persuasion-to-action strategies head-to-head, plus a control condition. Each strategy was grounded in established psychological or political persuasion literature and operationalized as a condition-specific system prompt appended after a common preamble (full prompt text in SI Appendix, Section 3.8.4):

\begin{enumerate}
    \item \textbf{Emotional Activation:} Activating empathy and moral outrage through vivid imagery, perspective-taking, and injustice framing to create an affective state that compels action \cite{ng2025influence, batson2011altruism, pagano2007moral}.
    \item \textbf{Implementation Intentions:} Forming concrete ``if-then'' action plans that specify the when, where, and how of the target behaviour, with the aim of bridging the intention--behaviour gap \cite{armitage2001efficacy, sheeran2025planning, sheeran2016intention}.
    \item \textbf{Identity Labeling:} Constructing and reinforcing a self-concept of the participant as someone who translates values into action, thereby casting behavioural compliance as self-congruent \cite{bryan2011motivating, bryan2011helping}.
    \item \textbf{Commitment Escalation:} Guiding the participant through a ladder of increasingly specific verbal commitments, from shared values to stated intentions to explicit pledges, so that compliance is framed as the self-consistent response \cite{freedman1966compliance, costa2018walking, cialdini2004social}.
    \item \textbf{Anticipated Regret:} Leading participants to envision the emotional consequences of failing to act, leveraging anticipated regret as a motivational force toward behavioural compliance \cite{brewer2016anticipated, abraham2003acting, sandberg2008anticipated}.
    \item \textbf{Information -- Issue:} Providing novel factual information, evidence, and expert findings about the issue itself, building a logical, evidence-based case for the importance and urgency of the cause \cite{petty1986communication, coppock2022persuasion}.
    \item \textbf{Information -- Impact Efficacy:} Providing concrete evidence of how specific individual actions translate into measurable outcomes \cite{vanzomeren2008integrative, camilleri2019collective, cryder2013donor}.
    \item \textbf{Mega (All Combined):} Model is given descriptions of all of the above strategies, can adaptively choose to use any or none \cite{hackenburg2025levers}.
\end{enumerate}

Participants were additionally randomized across AI model (GPT-4.1, Claude Opus 4.6, Grok 4, Gemini 3.1 Pro). Six issues were retained from Study~1, two of which were already high-salience, and two new high-salience issues were added, identified via Gallup's Most Important Problem measure \cite{jenningswill2011MIP} drawing on the 2025 UK Census and a 2026 YouGov poll. The primary behavioural outcome was petition signing; as a secondary outcome, participants directed a \pounds0.25 charity allocation to the sponsoring organization, GiveWell, or neither. All participants then completed strategy-specific self-report items, presented in randomised order (SI Appendix, Section 3.4).

\begin{table}[tbp]
    \centering
    \caption{Political issues used in Studies 1 and 2, with associated real UK petitions and sponsoring organizations.}
    \label{tab:petitions}
    \small
    \begin{tabular}{p{0.45\textwidth}ll}
        \toprule
        \textbf{Petition} & \textbf{Organization} & \textbf{Study} \\
        \midrule
        Stop the risk of nuclear war & Campaign for Nuclear Disarmament & 1 \\
        \addlinespace
        Stop banks fuelling the climate crisis & ActionAid & 1 \\
        \addlinespace
        Seats in parliament should match how we vote & Electoral Reform Society & 1 \& 2 \\
        \addlinespace
        End suffering for mother pigs & Compassion in World Farming & 1 \& 2 \\
        \addlinespace
        Oppose Facial Recognition & Liberty & 1 \& 2 \\
        \addlinespace
        Lift the ban on work for people seeking asylum & Refugee Action & 1 \& 2 \\
        \addlinespace
        Urge UK governments to protect trees & Woodland Trust & 1 \& 2 \\
        \addlinespace
        Stop Rising Child Poverty -- Scrap the Two-Child Limit \& Lock in Support & Save the Children & 1 \& 2 \\
        \addlinespace
        Continuing care for children with disabilities & Contact a Family & 2 \\
        \addlinespace
        Reducing economic inequality in the UK & The Equality Trust & 2 \\
        
        \bottomrule
    \end{tabular}
\end{table}

\subsection{Experimental materials}
Both studies were administered through a custom web application that managed participant randomization and delivered real-time multi-turn conversations with frontier LLMs via the OpenRouter API. All model hyperparameters were left at provider defaults. The same platform was used for both studies with minor design-specific modifications. All treatment conversations shared a common system-prompt preamble instructing the LLM to persuade the participant to sign the assigned petition without revealing its persuasive intent; control conversations used the same interface but focused on a neutral, non-political topic (fluoride, recycling, antibiotics, or artificial sweeteners). Full prompt text for all conditions is provided in SI Appendix, Section 3.8.4.

\subsection{Statistical analysis}
\textit{Study 1.} For each of six outcome variables, we estimated the average treatment effect of AI conversation relative to control via linear mixed-effects models (linear probability models for binary and proportional outcomes), adjusting for standardised pretreatment petition support, organisation support, charity interest, and random intercepts for issue (plus bonus amount for the donation outcome). Treatment effects were estimated at both the binary level (any treatment vs. control) and the condition level (petition-information, organisation-information, random prompt vs. control). To assess mechanisms, we correlated condition-level treatment effects with mean self-reported learning. Heterogeneous effects were estimated by splitting participants into prior supporters (pretreatment agreement $>$ 5) and others.

\textit{Study 2.} We estimated linear mixed-effects models of the form
\begin{equation}
\begin{split}
\text{DV} \sim{} & \text{strategy} + \text{scale}(\text{pre\_petition\_support}) \\
& + \text{scale}(\text{pre\_org\_support}) + (1 \mid \text{issue})
\end{split}
\end{equation}

where strategy is a nine-level factor (control as reference), pretreatment covariates are standardized, and issue enters as a random intercept. Petition signing (binary) was modeled as a linear probability model; attitudinal outcomes were standardized change scores, (post $-$ pre) / SD(pre). Each strategy was contrasted against control (Benjamini$-$Hochberg correction at $\alpha = 0.05$). To test whether information provision differentially drives attitudes versus behaviour, we compared the estimated treatment effects of the Information$-$Issue and Information$-$Impact Efficacy strategies relative to the other strategies, separately for behavioural and attitudinal model fits. Heterogeneous effects by prior support were assessed by interacting a binary treatment indicator (and, separately, the nine-level strategy factor) with standardized pretreatment petition support, estimated separately per outcome. Robustness checks re-estimated the primary model with (i) AI model as a fixed effect, (ii) demographic covariates, (iii) logistic mixed-effects specification for binary outcomes, and (iv) a leave-one-issue-out procedure. Manipulation checks tested whether each strategy elevated its corresponding mechanism items relative to the pooled mean of the remaining seven treatment strategies (excluding control). All models were fit using the \texttt{lme4} and \texttt{emmeans} packages in R; we report point estimates with 95\% confidence intervals.

\section*{Data Availability}

All code, data, and replication materials for both studies, as well as the SI Appendix, are publicly available in a \href{https://github.com/kobihackenburg/AI-action-persuasion}{GitHub repository}. All analyses were conducted in R (version $\geq$ 4.0).

\begin{ack}
None.
\end{ack}

\section*{Author Contributions}
K.H., L.H., C.W., C.S., and B.M.T. designed research; K.H., L.H., and C.W. performed research; K.H., L.H., and C.W. analyzed data; and K.H., L.H., C.W., C.S., and B.M.T. wrote the paper.

\noindent The authors declare no competing interests.

\bibliographystyle{plain}
\bibliography{references}

@article{luciano2024hypersuasion,
  title={Hypersuasion -- On AI's Persuasive Power and How to Deal with It},
  author={Luciano, F.},
  journal={Philosophy \& Technology},
  volume={37},
  pages={64},
  year={2024}
}

@misc{burtell2023artificial,
  title={Artificial Influence: An Analysis Of AI-Driven Persuasion},
  author={Burtell, M. and Woodside, T.},
  year={2023},
  note={arXiv:2303.08721 [Preprint]},
  doi={10.48550/arXiv.2303.08721}
}

@misc{chen2025framework,
  title={A Framework to Assess the Persuasion Risks Large Language Model Chatbots Pose to Democratic Societies},
  author={Chen, Z. and Kalla, J. and Le, Q. and Nakamura-Sakai, S. and Sekhon, J. and Wang, R.},
  year={2025},
  note={arXiv:2505.00036 [Preprint]},
  doi={10.48550/arXiv.2505.00036}
}

@misc{elsayed2024mechanism,
  title={A Mechanism-Based Approach to Mitigating Harms from Persuasive Generative AI},
  author={El-Sayed, S. and Akbulut, C. and McCroskery, A. and Keeling, G. and Kenton, Z. and Jalan, Z. and Marchal, N. and Manzini, A. and Shevlane, T. and Vallor, S. and Susser, D. and Franklin, M. and Bridgers, S. and Law, H. and Rahtz, M. and Shanahan, M. and Tessler, M. H. and Douillard, A. and Everitt, T. and Brown, S.},
  year={2024},
  note={arXiv:2404.15058 [Preprint]},
  doi={10.48550/arXiv.2404.15058}
}

@article{summerfield2024democracy,
  title={The impact of advanced AI systems on democracy},
  author={Summerfield, Christopher and Argyle, Lisa P and Bakker, Michiel and Collins, Teddy and Durmus, Esin and Eloundou, Tyna and Gabriel, Iason and Ganguli, Deep and Hackenburg, Kobi and Hadfield, Gillian K and others},
  journal={Nature Human Behaviour},
  pages={1--11},
  year={2025},
  publisher={Nature Publishing Group UK London}
}

@article{hackenburg2025levers,
  title={The levers of political persuasion with conversational artificial intelligence},
  author={Hackenburg, Kobi and Tappin, Ben M and Hewitt, Luke and Saunders, Ed and Black, Sid and Lin, Hause and Fist, Catherine and Margetts, Helen and Rand, David G and Summerfield, Christopher},
  journal={Science},
  volume={390},
  number={6777},
  pages={eaea3884},
  year={2025},
  publisher={American Association for the Advancement of Science}
}

@article{goldstein2024persuasive,
  title={How persuasive is AI-generated propaganda?},
  author={Goldstein, J. A. and Chao, J. and Grossman, S. and Stamos, A. and Tomz, M.},
  journal={PNAS Nexus},
  volume={3},
  pages={pgae034},
  year={2024}
}

@article{bai2023artificial,
  title={LLM-generated messages can persuade humans on policy issues},
  author={Bai, Hui and Voelkel, Jan G and Muldowney, Shane and Eichstaedt, Johannes C and Willer, Robb},
  journal={Nature Communications},
  volume={16},
  number={1},
  pages={6037},
  year={2025},
  publisher={Nature Publishing Group UK London}
}

@article{costello2024durably,
  title={Durably reducing conspiracy beliefs through dialogues with AI},
  author={Costello, T. H. and Pennycook, G. and Rand, D. G.},
  journal={Science},
  volume={385},
  pages={eadq1814},
  year={2024}
}

@article{hackenburg2025scaling,
  title={Scaling language model size yields diminishing returns for single-message political persuasion},
  author={Hackenburg, K. and Tappin, B. M. and R{\"o}ttger, P. and Hale, S. A. and Bright, J. and Margetts, H.},
  journal={Proceedings of the National Academy of Sciences},
  volume={122},
  pages={e2413443122},
  year={2025}
}

@misc{schoenegger2025llms,
  title={Large Language Models Are More Persuasive Than Incentivized Human Persuaders},
  author={Schoenegger, P. and Salvi, F. and Liu, J. and Nan, X. and Debnath, R. and Fasolo, B. and Leivada, E. and Recchia, G. and G{\"u}nther, F. and Zarifhonarvar, A. and Kwon, J. and Islam, Z. U. and Dehnert, M. and Lee, D. Y. H. and Reinecke, M. G. and Kamper, D. G. and Koba{\c{s}}, M. and Sandford, A. and Kgomo, J. and Hewitt, L. and Kapoor, S. and Oktar, K. and Kucuk, E. E. and Feng, B. and Jones, C. R. and Gainsburg, I. and Olschewski, S. and Heinzelmann, N. and Cruz, F. and Tappin, B. M. and Ma, T. and Park, P. S. and Onyonka, R. and Hjorth, A. and Slattery, P. and Zeng, Q. and Finke, L. and Grossmann, I. and Salatiello, A. and Karger, E.},
  year={2025},
  note={arXiv:2505.09662 [Preprint]},
  doi={10.48550/arXiv.2505.09662}
}

@article{simchon2024persuasive,
  title={The persuasive effects of political microtargeting in the age of generative artificial intelligence},
  author={Simchon, A. and Edwards, M. and Lewandowsky, S.},
  journal={PNAS Nexus},
  volume={3},
  pages={pgae035},
  year={2024}
}

@article{argyle2025testing,
  title={Testing theories of political persuasion using AI},
  author={Argyle, L. P. and Busby, E. C. and Gubler, J. R. and Lyman, A. and Olcott, J. and Pond, J. and Wingate, D.},
  journal={Proceedings of the National Academy of Sciences},
  volume={122},
  pages={e2412815122},
  year={2025}
}

@article{hainmueller2015validating,
  title={Validating vignette and conjoint survey experiments against real-world behavior},
  author={Hainmueller, J. and Hangartner, D. and Yamamoto, T.},
  journal={Proceedings of the National Academy of Sciences},
  volume={112},
  pages={2395--2400},
  year={2015}
}

@article{okeefe2021persuasive,
  title={Persuasive Message Pretesting Using Non-Behavioral Outcomes: Differences in Attitudinal and Intention Effects as Diagnostic of Differences in Behavioral Effects},
  author={O'Keefe, D. J.},
  journal={Journal of Communication},
  volume={71},
  pages={623--645},
  year={2021}
}

@article{coppock2015assessing,
  title={Assessing the Correspondence between Experimental Results Obtained in the Lab and Field: A Review of Recent Social Science Research},
  author={Coppock, A. and Green, D. P.},
  journal={Political Science Research and Methods},
  volume={3},
  pages={113--131},
  year={2015}
}

@article{saccardo2024field,
  title={Field testing the transferability of behavioural science knowledge on promoting vaccinations},
  author={Saccardo, S. and Dai, H. and Han, M. A. and Vangala, S. and Hoo, J. and Fujimoto, J.},
  journal={Nature Human Behaviour},
  volume={8},
  pages={878--890},
  year={2024}
}

@article{dellavigna2018motivates,
  title={What Motivates Effort? Evidence and Expert Forecasts},
  author={DellaVigna, S. and Pope, D.},
  journal={Review of Economic Studies},
  volume={85},
  pages={1029--1069},
  year={2018}
}

@article{gerber2008social,
  title={Social Pressure and Voter Turnout: Evidence from a Large-Scale Field Experiment},
  author={Gerber, A. S. and Green, D. P. and Larimer, C. W.},
  journal={American Political Science Review},
  volume={102},
  pages={33--48},
  year={2008}
}

@article{ng2025influence,
author = {Ng, Wei Jie Reiner and See, Ya Hui Michelle and Cheung, Mike W -L},
copyright = {Copyright 2025 Elsevier B.V., All rights reserved.},
issn = {0021-9916},
journal = {Journal of communication},
language = {eng},
number = {2},
pages = {101-111},
title = {The Influence of affective and cognitive appeals on persuasion outcomes: a cross-cultural meta-analysis},
volume = {75},
year = {2025},
}

@book{batson2011altruism,
author = {Batson, C. Daniel},
address = {New York},
copyright = {Copyright 2015 Elsevier B.V., All rights reserved.},
edition = {1},
isbn = {9780195341065},
language = {eng},
publisher = {Oxford University Press},
title = {Altruism in humans},
year = {2011},
}

@article{pagano2007moral,
author = {Pagano, Sabrina J. and Huo, Yuen J.},
address = {Malden, USA},
copyright = {Copyright 2007 International Society of Political Psychology},
issn = {0162-895X},
journal = {Political psychology},
language = {eng},
number = {2},
pages = {227-255},
publisher = {Blackwell Publishing Inc},
title = {The Role of Moral Emotions in Predicting Support for Political Actions in Post-War Iraq},
volume = {28},
year = {2007},
}

@article{armitage2001efficacy,
author = {Armitage, Christopher J. and Conner, Mark},
address = {Oxford, UK},
copyright = {2001 The British Psychological Society},
issn = {0144-6665},
journal = {British journal of social psychology},
language = {eng},
number = {4},
pages = {471-499},
publisher = {Blackwell Publishing Ltd},
title = {Efficacy of the Theory of Planned Behaviour: A meta-analytic review},
volume = {40},
year = {2001},
}

@article{sheeran2025planning,
author = {Sheeran, Paschal and Listrom, Olivia and Gollwitzer, Peter M.},
address = {Abingdon},
copyright = {2024 European Association of Social Psychology 2024},
issn = {1046-3283},
journal = {European review of social psychology},
language = {eng},
number = {1},
pages = {162-194},
publisher = {Routledge},
title = {The when and how of planning: Meta-analysis of the scope and components of implementation intentions in 642 tests},
volume = {36},
year = {2025},
}

@article{sheeran2016intention,
author = {Sheeran, Paschal and Webb, Thomas L.},
copyright = {2016 John Wiley & Sons Ltd},
issn = {1751-9004},
journal = {Social and personality psychology compass},
language = {eng},
number = {9},
pages = {503-518},
publisher = {Blackwell Publishing Ltd},
title = {The Intention-Behavior Gap},
volume = {10},
year = {2016},
}

@article{bryan2011motivating,
author = {Bryan, Christopher J and Walton, Gregory M and Rogers, Todd and Dweck, Carol S},
address = {United States},
copyright = {copyright © 1993–2008 by the National Academy of Sciences of the United States of America},
issn = {0027-8424},
journal = {Proceedings of the National Academy of Sciences - PNAS},
language = {eng},
number = {31},
pages = {12653-12656},
publisher = {National Academy of Sciences},
title = {Motivating voter turnout by invoking the self},
volume = {108},
year = {2011},
}

@article{bryan2011helping,
author = {Bryan, Christopher J. and Master, Allison and Walton, Gregory M.},
address = {United States},
copyright = {Child Development © 2014 Society for Research in Child Development, Inc.},
issn = {0009-3920},
journal = {Child development},
language = {eng},
number = {5},
pages = {1836-1842},
publisher = {Blackwell Publishing Ltd},
title = {"Helping" Versus "Being a Helper": Invoking the Self to Increase Helping in Young Children},
volume = {85},
year = {2014},
}

@article{freedman1966compliance,
author = {Freedman, Jonathan L and Fraser, Scott C},
address = {United States},
copyright = {1966 American Psychological Association},
issn = {0022-3514},
journal = {Journal of personality and social psychology},
language = {eng},
number = {2},
pages = {195-202},
publisher = {American Psychological Association},
title = {Compliance without pressure: The foot-in-the-door technique},
volume = {4},
year = {1966},
}

@article{costa2018walking,
author = {Costa, Mia and Schaffner, Brian F. and Prevost, Alicia},
address = {United States},
copyright = {Copyright 2018 Elsevier B.V., All rights reserved.},
issn = {1932-6203},
journal = {PloS one},
language = {eng},
number = {5},
pages = {e0197066-},
publisher = {Public Library of Science},
title = {Walking the walk? Experiments on the effect of pledging to vote on youth turnout},
volume = {13},
year = {2018},
}

@article{cialdini2004social,
author = {Cialdini, Robert B. and Goldstein, Noah J.},
address = {Palo Alto, CA},
copyright = {Copyright 2011 Elsevier B.V., All rights reserved.},
issn = {0066-4308},
journal = {Annual review of psychology},
language = {eng},
number = {1},
pages = {591-621},
publisher = {Annual Reviews},
title = {Social Influence: Compliance and Conformity},
volume = {55},
year = {2004},
}

@article{brewer2016anticipated,
author = {Brewer, Noel T. and DeFrank, Jessica T. and Gilkey, Melissa B.},
address = {United States},
copyright = {2016 American Psychological Association},
issn = {0278-6133},
journal = {Health psychology},
language = {eng},
number = {11},
pages = {1264-1275},
publisher = {American Psychological Association},
title = {Anticipated Regret and Health Behavior: A Meta-Analysis},
volume = {35},
year = {2016},
}

@article{abraham2003acting,
author = {Abraham, Charles and Sheeran, Paschal},
address = {Oxford, UK},
copyright = {2003 The British Psychological Society},
issn = {0144-6665},
journal = {British journal of social psychology},
language = {eng},
number = {4},
pages = {495-511},
publisher = {Blackwell Publishing Ltd},
title = {Acting on intentions: The role of anticipated regret},
volume = {42},
year = {2003},
}

@article{sandberg2008anticipated,
author = {Sandberg, Tracy and Conner, Mark},
address = {Oxford, UK},
copyright = {2008 The British Psychological Society},
issn = {0144-6665},
journal = {British journal of social psychology},
language = {eng},
number = {4},
pages = {589-606},
publisher = {Blackwell Publishing Ltd},
title = {Anticipated regret as an additional predictor in the theory of planned behaviour: A meta-analysis},
volume = {47},
year = {2008},
}

@article{vanzomeren2008integrative,
author = {van Zomeren, Martijn and Postmes, Tom and Spears, Russell},
address = {Washington, DC},
copyright = {2008 American Psychological Association},
issn = {0033-2909},
journal = {Psychological bulletin},
language = {eng},
number = {4},
pages = {504-535},
publisher = {American Psychological Association},
title = {Toward an Integrative Social Identity Model of Collective Action: A Quantitative Research Synthesis of Three Socio-Psychological Perspectives},
volume = {134},
year = {2008},
}

@article{camilleri2019collective,
author = {Camilleri, Adrian R. and Larrick, Richard P.},
address = {United States},
copyright = {2019 American Psychological Association},
issn = {0096-3445},
journal = {Journal of experimental psychology. General},
language = {eng},
number = {3},
pages = {550-569},
publisher = {American Psychological Association},
title = {The Collective Aggregation Effect: Aggregating Potential Collective Action Increases Prosocial Behavior},
volume = {148},
year = {2019},
}

@article{cryder2013donor,
author = {Cryder, Cynthia E. and Loewenstein, George and Scheines, Richard},
address = {Kidlington},
copyright = {2012 Elsevier Inc.},
issn = {0749-5978},
journal = {Organizational behavior and human decision processes},
language = {eng},
number = {1},
pages = {15-23},
publisher = {Elsevier Inc},
title = {The donor is in the details},
volume = {120},
year = {2013},
}

@book{petty1986communication,
author = {Petty, Richard E and Cacioppo, John T},
address = {New York, NY},
copyright = {Springer-Verlag New York 1986},
edition = {1},
isbn = {1461293782},
language = {eng},
organization = {SpringerLink (Online service)},
publisher = {Springer New York},
series = {Springer Series in Social Psychology},
title = {Communication and Persuasion: Central and Peripheral Routes to Attitude Change},
year = {1986},
}

@book{coppock2022persuasion,
author = {Coppock, Alexander},
address = {Chicago},
edition = {1},
isbn = {0226821846},
language = {eng},
publisher = {The University of Chicago Press},
series = {Chicago Studies in American Politics},
title = {Persuasion in parallel : how information changes minds about politics},
year = {2022},
}

@misc{tappin2025exposure,
  title={For Digital Mass Persuasion, Exposure Matters More Than Persuasiveness},
  author={Tappin, Ben M.},
  year={2025},
  month={December},
  doi={10.31234/osf.io/5ewbu_v1},
  url={https://osf.io/preprints/psyarxiv/5ewbu_v1},
  publisher={PsyArXiv},
  note={Preprint}
}

@article{Paler2018socialcosts,
  title={The Social Costs of Public Political Participation: Evidence from a Petition Experiment in Lebanon},
  author={Laura Paler and Leslie Marshall and Sami Atallah},
  journal={The Journal of Politics},
  year={2018},
  volume={80},
  pages={1405 - 1410},
  url={https://api.semanticscholar.org/CorpusID:158621307}
}

@article{Coppock2016field,
author = {Coppock, Alexander and Guess, Andrew and Ternovski, John},
address = {New York},
copyright = {Springer Science+Business Media New York 2015},
issn = {0190-9320},
journal = {Political behavior},
language = {eng},
number = {1},
pages = {105-128},
publisher = {Springer Science + Business Media},
title = {When Treatments are Tweets: A Network Mobilization Experiment over Twitter},
volume = {38},
year = {2016},
}

@article{jenningswill2011MIP,
author = {Jennings, Will and Wlezien, Christopher},
address = {Oxford},
copyright = {Copyright © 2011 American Association for Opinion Research},
issn = {0033-362X},
journal = {Public opinion quarterly},
language = {eng},
number = {3},
pages = {545-555},
publisher = {Oxford University Press},
title = {DISTINGUISHING BETWEEN MOST IMPORTANT PROBLEMS AND ISSUES?},
volume = {75},
year = {2011},
}

@article{peereyal2022prolific,
author = {Peer, Eyal and Rothschild, David and Gordon, Andrew and Evernden, Zak and Damer, Ekaterina},
address = {New York},
copyright = {The Psychonomic Society, Inc. 2021},
issn = {1554-3528},
journal = {Behavior research methods},
language = {eng},
number = {4},
pages = {1643-1662},
publisher = {Springer US},
title = {Data quality of platforms and panels for online behavioral research},
volume = {54},
year = {2022},
}

@misc{stagnaro2025prolific, 
title={Representativeness and Response Validity Across Nine Opt-In Online Samples}, 
url={osf.io/preprints/psyarxiv/h9j2d_v2}, 
DOI={10.31234/osf.io/h9j2d_v2},
publisher={PsyArXiv}, 
author={Stagnaro, Michael N and Druckman, James and Berinsky, Adam J and Arechar, Antonio A and Willer, Robb and Rand, David G}, 
year={2025}, 
month={Jul} 
}

@misc{akbulut2026gdm,
      title={Evaluating Language Models for Harmful Manipulation}, 
      author={Canfer Akbulut and Rasmi Elasmar and Abhishek Roy and Anthony Payne and Priyanka Suresh and Lujain Ibrahim and Seliem El-Sayed and Charvi Rastogi and Ashyana Kachra and Will Hawkins and Kristian Lum and Laura Weidinger},
      year={2026},
      eprint={2603.25326},
      archivePrefix={arXiv},
      primaryClass={cs.AI},
      url={https://arxiv.org/abs/2603.25326}, 
}

\end{document}